\input harvmac
\noblackbox
\def\pano{\par\noindent}

\def\bigno{\bigskip\noindent}
\font\cmss=cmss10
\font\cmsss=cmss10 at 7pt
\def\rlx{\relax\leavevmode}
\def\inbar{\vrule height1.5ex width.4pt depth0pt}
\def\IC{\relax\,\hbox{$\inbar\kern-.3em{\rm C}$}}
\def\IR{\relax{\rm I\kern-.18em R}}
\def\IN{\relax{\rm I\kern-.18em N}}
\def\IP{\relax{\rm I\kern-.18em P}}
\def\ZZ{\rlx\leavevmode\ifmmode\mathchoice{\hbox{\cmss Z\kern-.4em Z}}
 {\hbox{\cmss Z\kern-.4em Z}}{\lower.9pt\hbox{\cmsss Z\kern-.36em Z}}
 {\lower1.2pt\hbox{\cmsss Z\kern-.36em Z}}\else{\cmss Z\kern-.4em Z}\fi}
\def\narrowplus{\kern -.04truein + \kern -.03truein}
\def\narrowminus{- \kern -.04truein}
\def\narrowminussub{\kern -.02truein - \kern -.01truein}

\def\cl{\centerline}

\def\oo#1{\bar{#1}}

\def\wt#1{\widetilde{#1}}

\def\type{type$\,{\rm I}\ $}
\def\Type{Type$\,{\rm I}\ $}
\def\typea{type$\,{\rm II} $}
\def\typeb{type$\,{\rm II}\,{\rm B}\ $}
\def\Typeb{Type$\,{\rm II}\,{\rm B}\ $}

\def\sqr#1#2{{\vcenter{\vbox{\hrule height.#2pt
 \hbox{\vrule width.#2pt height#1pt \kern#1pt
 \vrule width.#2pt}\hrule height.#2pt}}}}


\lref\rbback{M.\ Bianchi, G.\ Pradisi and A.\ Sagnotti, {\it Toroidal
Compactifications and Symmetry Breaking in Open String Theories},
Nucl.\ Phys.\ {\bf B376} (1992) 365\semi
Z.\ Kakushadze, G.\ Shiu and S.-H.H.\ Tye,
{\it Type IIB Orientifolds
with NS-NS Antisymmetric Tensor Backgrounds}, hep-th/9803141}

\lref\rbisag{
M.\ Bianchi and A.\ Sagnotti, {\it On the Systematics of Open String
Theories}, Phys.\ Lett.\ {\bf B247} (1990) 517;
{\it Twist Symmetry and Open String Wilson Lines},
Nucl.\ Phys.\ {\bf B361} (1991) 519}

\lref\rcardy{J.L.\ Cardy, {Boundary Conditions, Fusion Rules and the
Verlinde Formula}, Nucl.\ Phys.\ {\bf B324} (1989) 581}

\lref\ritalians{G.\ Pradisi, A.\ Sagnotti and Y.S.\ Stanev, {\it Planar
Duality in $SU(2)$ WZW Models}, Phys.\ Lett.\ {\bf B354} (1995) 279,
hep-th/9503207;
{\it The Open Descendants of Non-Diagonal $SU(2)$ WZW Models},
Phys.\ Lett.\ {\bf B356} (1995) 230, hep-th/9506014;
{\it Completeness Conditions for Boundary Operators in $2D$ Conformal
Field Theory}, Phys.\ Lett.\ {\bf B381} (1996) 97, hep-th/9603097 \semi
J. Fuchs and C. Schweigert, {\it Branes:From Free Fields to General
Backgrounds}, hep-th/9712257}

\lref\rgep{D.\ Gepner, {\it Space-time Supersymmetry in Compactified
String Theory and Superconformal Models}, Nucl.\ Phys.\ {\bf B296} (1988) 757}

\lref\rfss{J.\ Fuchs, A.N.\ Schellekens and C.\ Schweigert, {\it A Matrix S
for all Simple Current Extensions}, Nucl.\ Phys.\ {\bf B473} (1996) 323,
hep-th/960078}

\lref\rgepner{C.\ Angelantonj, M.\ Bianchi, G.\ Pradisi, A.\ Sagnotti and
Ya.S.\ Stanev, {\it Comments on Gepner Models and Type I Vacua in String
Theory}, Phys.\ Lett.\ {\bf B387} (1996) 743, hep-th/9607229}

\lref\rfourdim{
M.\ Berkooz and R.G.\ Leigh, {\it A $D=4$ $N=1$ Orbifold of Type I Strings},
Nucl.\ Phys.\ {\bf B483} (1997) 187, hep-th/9605049\semi
C.\ Angelantonj, M.\ Bianchi, G.\ Pradisi, A.\ Sagnotti and
Ya.S.\ Stanev, {\it Chiral Asymmetry in Four-Dimensional Open-String Vacua},
Phys.\ Lett.\ {\bf B385} (1996) 96, hep-th/9606169\semi
Z.\ Kakushadze, {\it Aspects of $N=1$ Type I-Heterotic Duality in
 Four Dimensions}, Nucl.\ Phys.\ {\bf B512} (1998) 221,
hep-th/9704059\semi
Z.\ Kakushadze and G.\ Shiu, {\it A Chiral $N=1$ Type I Vacuum in Four
Dimensions and Its Heterotic Dual}, Phys.\ Rev.\ {\bf D56} (1997) 3686,
hep-th/9705163;
{\it $4D$ Chiral $N=1$ Type I Vacua with and
without $D5$-branes}, Nucl.\ Phys.\ {\bf B520} (1998) 75, hep-th/9706051\semi
Z.\ Kakushadze, {\it A Three-Family $SU(6)$ Type I Compactification},
hep-th/9804110;
{\it A Three-Family $SU(4)_c\otimes SU(2)_w\otimes U(1)$ Type I
Vacuum}, hep-th/9806044\semi
G.\ Zwart, {\it Four-dimensional $N=1$ $\ZZ_N \times \ZZ_M$ Orientifolds},
hep-th/9708040\semi
G.\ Aldazabal, A.\ Font, L.E.\ Ib{\'a}{\~n}ez and G.\ Violero,
{\it $D=4$, $N=1$, Type IIB Orientifolds}, hep-th/9804026}

\lref\rkst{
Z.\ Kakushadze, G.\ Shiu and S.-H.H.\ Tye, {\it Type IIB Orientifolds,
F-theory, Type I Strings on Orbifolds and Type I-Heterotic Duality},
hep-th/9804092\semi
G.\ Shiu and S.-H.H.\ Tye, {\it TeV Scale Superstring and Extra Dimensions},
hep-th/9805157\semi
Z.\ Kakushadze, {\it On Four Dimensional $N=1$ Type I Compactifications},
hep-th/9806008}

\lref\rpolwit{J.\ Polchinski and E.\ Witten, {\it Evidence for
Heterotic-\type String Duality}, Nucl.\ Phys.\ {\bf B460} (1996) 525,
hep-th/9510169}

\lref\rgimpol{
E.G.\ Gimon and J.\ Polchinski, {\it Consistency Conditions
for Orientifolds and D-Manifolds}, Phys.\ Rev.\ {\bf D54} (1996) 1667,
hep-th/9601038}

\lref\rsechs{
E.G.\ Gimon and C.V.\ Johnson, {\it $K3$ Orientifolds},
Nucl.\ Phys.\ {\bf B477} (1996) 715, hep-th/9604129\semi
A.\ Dabholkar and J.\ Park, {\it Strings on Orientifolds}, Nucl.\ Phys.
{\bf B477} (1996) 701}

\lref\rbsf{R.\ Blumenhagen, R.\ Schimmrigk and A.\ Wi{\ss}kirchen,
 {\it $(0,2)$ mirror symmetry},
 Nucl.\ Phys.\ {\bf B486} (1997) 598, hep-th/9609167\semi
 R.\ Blumenhagen and S.\ Sethi,
 {\it On orbifolds of $(0,2)$ models},
 Nucl.\ Phys.\ {\bf B491} (1997) 263, hep-th/9611172\semi
 R.\ Blumenhagen and M.\ Flohr,
 {\it Aspects of $(0,2)$ orbifolds and mirror symmetry},
 Phys.\ Lett.\ {\bf B404} (1997) 41, hep-th/9702199\semi
 R.\ Blumenhagen, {\it Target Space Duality for $(0,2)$ compactifications}
 Nucl.\ Phys.\ {\bf B513} (1998) 573, hep-th/9707198;
 {\it $(0,2)$ Target Space Duality, CICYs and Reflexive
 Sheaves}, Nucl.\ Phys.\ {\bf B514} (1998) 688, hep-th/9710021\semi
 A.\ Sen and S.\ Sethi, {\it the Mirror Transform of \type Vacua in Six
 Dimensions}, Nucl.\ Phys.\ {\bf B499} (1997) 573, hep-th/9703157 }

\lref\rfseins{J.\ Fuchs and C.\ Schweigert,
{\it Classifying Algebras for Boundary Conditions and Traces on Spaces
of Conformal Blocks}, hep-th/9801191}
\lref\rfszwei{J.\ Fuchs and C.\ Schweigert,
{\it A Classifying Algebra for Boundary Conditions},
Phys.\ Lett.\ {\bf B414} (1997) 251, hep-th/9708141}

\lref\randy{A.\ Recknagel and V.\ Schomerus,
{\it D-branes in Gepner Models}, hep-th/9712186}

\Title{\vbox{\hbox{hep--th/9806131}
 \hbox{IASSNS--HEP--98/54}
 \hbox{BONN--TH--98--12}}}
{\vbox{ \hbox{Spectra of 4D, N=1 \Type String Vacua} \vskip 0.5cm
 \hbox{\phantom{Spe} on Non-Toroidal CY Threefolds }}}
\smallskip
\centerline{Ralph Blumenhagen${}^1$ and Andreas Wi\ss kirchen${}^2$}
\bigskip
\centerline{${}^{1}$ \it School of Natural Sciences,
 Institute for Advanced Study,}
\centerline{\it Olden Lane, Princeton NJ 08540, USA}
\smallskip
\centerline{${}^{2}$ \it Physikalisches Institut der Universit\"at Bonn,}
\centerline{\it Nu{\ss}allee 12, 53115 Bonn, Germany}

\smallskip
\bigskip
\bigskip\bigskip
\centerline{\bf Abstract}
\noindent
We compute the massless spectra of some four dimensional, $N=1$
supersymmetric compactifications of the \type string.
The backgrounds are non-toroidal Calabi-Yau manifolds described at special
points in moduli space by Gepner models.
Surprisingly, the abstract conformal field theory computation reveals
Chan-Paton gauge groups as big as $SO(12)\otimes SO(20)$ or
$SO(8)^4\otimes SO(4)^3$.

\footnote{}
{\pano
${}^1$ e--mail:\ blumenha@sns.ias.edu
\pano
${}^2$ e--mail:\ wisskirc@avzw02.physik.uni--bonn.de
\pano}
\Date{06/98}
\newsec{Introduction}

Of all five ten dimensional string theories the \type branch
has certainly received less attention during the perturbative
decade in the history of this field. However, mainly during the last
two years we have seen some extended activity in computing and better
understanding \type vacua in six and four space time dimensions
\refs{\rbisag\rgimpol\rsechs{--}\rfourdim}. Technically, this progress
was fed by our better understanding of orientifolds and D-branes.
Moreover, we know that all five string theories are related to
each other by dualities and are believed to have a common ancestor in eleven
dimensions, so called M-theory. From this point of view every
string theory can teach us something about M-theory.

 From the low energy effective actions of the \type and the $SO(32)$ heterotic
string, one conjectures a strong-weak coupling duality between these two
theories \rpolwit. This duality has passed several test, even after
compactification. In particular, in four dimensions one expects a purely
perturbative duality between these two string theories in special regions of
the moduli space. For some orientifolds there also exists
a duality to a certain kind of F-theory vacua, which can be used
to gain some non-perturbative insight into the former models.

So far all \type model building was restricted to toroidal orbifolds
in general including 9-branes and 5-branes for tadpole cancellation.
As argued in \rkst\ consistency of such orientifolds severely constraints
the number of such compactifications. For instance, absence of additional
non-perturbative states and world sheet consistency allows only six
possible abelian orbifold of $T^6$.

This paper deals with the construction of four dimensional \type vacua
for background manifolds which are not toroidal orbifolds. Since
the exact conformal field theory for every point in moduli space is not known,
the very illustrative method of \rgimpol\ can not be used. However,
for some Calabi-Yau threefolds Gepner has provided exactly solvable points
in terms of tensor products of $N=2$ superconformal field theories \rgep.
Furthermore, in a series of papers \ritalians, purely algebraic methods to
 construct \type descendants of such abstract conformal field theories were
developed\footnote{$^1$}{We would like to emphasize that already in 1991
what is now called the Gimon-Polchinski model was discussed in a paper by
Bianchi and Sagnotti \rbisag.}. The open string sector of Gepner models
was also discussed in \randy.

After reviewing the general construction of \type vacua in terms of
rational conformal field theories in section two, in section three
we apply this to the case of Gepner models in four dimensions.
 From a naive approach such models
are hard to come by, for their huge number of representations of the
chiral algebra. We make use of simple current techniques to first transform
the model to accessible size. In particular we consider a $\ZZ_5$ orbifold
of the quintic and a $\ZZ_3$ orbifold of $\IP_{1,1,1,3,3}[9]$.
The \type models turn out to have interesting spectra with gauge groups
like for instance $SO(12)\times SO(20)$.
Unfortunately, in this formalism there is no inherent distinction between
Chan-Paton (CP) factors arising from 9-branes and 5-branes, so that
in contrast to toroidal orbifolds it is more involved to find
dual heterotic partners. Since the threefolds we consider do not admit an
elliptic fibration, there does not exist an F-theory dual model, either.

\newsec{\Type descendants of \Typeb vacua }

For constructing vacua of closed string theories, one loop modular
invariance served as the main constraint in ruling out non consistent
compactifications. Fortunately,
for open string vacua, i.e.\ for conformal field theories (CFTs) with
boundaries, one loop tadpole cancellation turned out to be as restrictive.
In this section we briefly review the essential steps for
constructing all four \type one loop amplitudes in terms of a
rational conformal field theory. For further details we refer the reader to
the extensive literature \ritalians.

One starts with a left right symmetric \typeb model with a set of
representations of the maximal
chiral algebra ${\cal A}$ with characters $\chi_i$. These $\chi_i$ transform
under the two generating modular transformations with some matrices $S$ and
$T$. The charge conjugation matrix is defined as $C=S^2$. The
modular invariant torus partition function can be written as
\eqn\torus{ Z_T(\tau,\oo\tau)=\sum_{i,j=1}^{N} \chi_i(\tau)\ N_{ij}\
 \oo\chi_j(\oo\tau).}
Since one started with a left-right symmetric CFT, the world sheet
parity operation $\Omega$ is really a symmetry of the two dimensional
world sheet theory and can thus be divided out\footnote{$^2$}{Note, that
for toroidal orbifolds, requiring that $\Omega$ is really a symmetry
of the theory yields further severe conditions arising from the twisted sector
two form flux \rkst. Thus, the world sheet consistency conditions discussed
in \rkst\ seem automatically to be satisfied in the rational CFT approach.
Whether this excludes any subtleties related to the appearance of additional
non-perturbative states is not absolutely clear to us.}.
As usual, modding out $\Omega$ introduces
unoriented world sheets like $\IR\IP_2$ at tree level or the Klein bottle
at one loop, respectively.

In the direct (open channel) Klein bottle amplitude only left-right
symmetric states are allowed to contribute, leading to
\eqn\klein{ {K}(y)= \sum_{j=1}^{N} K_j\ \chi_j(2iy),}
where the variable $y$ is real. The coefficients have to satisfy
$|K_j|=N_{jj}$ and the choice of signs has to be consistent with
the fusion rules. In the course of this paper, we will always choose the
standard Klein bottle projection with all signs positive, implying
(anti)symmetrization in the (R-R) NS-NS closed string sector.
The transverse (closed channel) amplitude is obtained by applying the
modular S-matrix to the direct Klein bottle amplitude.
Since roughly speaking the Klein bottle can be written as
a cylinder with two crosscaps and the closed channel describes a closed string
propagating between these two boundaries, the transverse amplitude
can be written as
\eqn\kleint{ \wt{K}(t)= \sum_{j=1}^{N} \Gamma_j^2\ \chi_j(it/2).}
Integrating \kleint\ over $t$, in general
leads to infinities interpreted as massless tadpoles, which have to be
cancelled by other contributions arising from open string diagrams.

The open channel cylinder amplitude can be written as
\eqn\cyl{ A(y)=\sum_{j=1}^{N} \sum_{a,b} A_{ab}^j\ n^a n^b\
 \chi_j(iy/2),}
where the $n^a$ denote Chan-Paton factors and the coefficient $A_{ab}^i$ are
non-negative integers.
Since the transverse amplitude describes the propagation of a closed
string between two boundaries (D-branes), it can again be written as a square
\eqn\cylt{ \wt{A}(t)= \sum_{j=1}^{N} (\sum_a B_a^j n^a)^2 \
 \chi_j(2it).}
For trivial choice of the gluing automorphism a state in $\chi_j$ is
reflected into its charge conjugate at the
boundary. Thus, only states which couple to its conjugate in $Z_T$ are
allowed to contribute in the closed string channel \cylt.
As Cardy has shown, choosing the torus partition function to be
the charge conjugation one and taking the same number of CP factors
than characters, a natural solution is
$A_{ij}^k=N_{ij}^k$ \rcardy. Moreover, in the closed string channel
one obtains
\eqn\cylt{ B_i^j={S_{ij}\over \sqrt{S_{1j}}}.}
Finally, the M\"obius strip also contributes at one loop in the open string
sector
\eqn\moebius{ M(y)=\sum_j M_{a}^j\ n^a\ \hat{\chi}_j
 \left({iy+1\over 2}\right),}
where it is convenient to introduce the real characters
\eqn\rchar{ \hat{\chi}\left({iy+1\over 2}\right)=e^{-i\pi
 \left(h-{c\over 24}\right)} {\chi}\left({iy+1\over 2}\right).}
The transverse M\"obius strip amplitude can be written as
\eqn\moebiust{ \wt{M}(t)=\sum_j \Gamma_j \left[\sum_a B_a^j n^a\right]
 \hat{\chi}_j\left({it+1\over 2}\right) }
reflecting the fact that the M\"obius strip can be regarded as a cylinder
with a crosscap on one side.
The transverse amplitude is related to the direct amplitude \moebius\
by an action of the matrix
$P=T^{1\over 2}ST^2ST^{1\over 2}$. In order to have a consistent projection
in the open sector the cylinder and M\"obius strip amplitude must satisfy
\eqn\relat{ M_a^j=A_{aa}^j\ {\rm mod}\ 2.}
As was shown in \ritalians, besides the fusion rule coefficients
 one can define another integer valued object \rfseins
\eqn\yps{ Y_{ij}^k=\sum_l {S_{il}\, P_{jl}\, P_{kl}^* \over S_{1l} }.}
For the charge conjugation modular invariant, $Y_{ij}^{k}$ determines a
consistent CP assignment by
\eqn\CPs{ K_j=Y_{j1}^1, \quad\quad M_{i}^j=Y_{ij}^1.}
In the following we will only make use of this choice of CP factors.
Summarizing, in order to compute \type one loop amplitudes one has
to specify a rational CFT including a set of characters, yielding a
representation of the modular group, and the matrices $S$, $T$ and $P$.

Suppose, a second modular invariant torus amplitude $Z_T'$ is related to
the charge conjugation one $Z_T^c$ by a simple current acting without
fixed points. Thus, every orbit has the same length $L$.
Denote by $M'$ the set of those $\chi_i$ coupling to its
charge dual in $Z_T'$.
Then, one can easily show that the two Klein bottle amplitudes are related by
\eqn\kbrel{ \wt{K}'_j=\cases{ L\ \wt{K}^c_j \quad & {\rm for}\
 $j\in M'$ \cr
 0 & {\rm for}\ $j\not\in M'$.\cr } }
If one identifies CP factors lying in the same orbit under the simple current
one can naturally define a new cylinder and M\"obius strip amplitude
\eqn\crel{\eqalign{ A'&={1\over L} A^c|_{{\rm identify}\;CPs},\cr
 M'&=M^c|_{{\rm identify}\;CPs} } }
automatically satisfying all the consistency
requirements and $A_j'=0$ for $j\not\in M'$.
The generalization of this procedure to simple currents with fixed
points was discussed in \refs{\ritalians,\rfszwei}.
For all cases studied in the sequel of this paper, the two formulae
\kbrel\ and \crel\ are however sufficient.

In the case of a four dimensional Gepner model with $N_f$ tensor factors
the characters $\chi_i$ are orbits under a set of simple current including
the GSO projection\footnote{$^3$}{We denote the primary field with $N=2$
quantum numbers $(l,m,s)$ by $\Phi^l_{m,s}$.}
\eqn\simple{ J=\left(\Phi^0_{1,1}\right)^{N_f}\otimes s_{SO(2)}, }
where $s_{SO(2)}$ denotes the spinor representation of $SO(2)$
arising from two transverse fermions which are superpartners
of the two bosonic space time coordinates $X^\mu$. Remember, as
q-series these orbits vanish due to supersymmetry.
The overall integrated transverse Klein bottle amplitude looks like
\eqn\kleinto{ \wt{K}=\int_0^\infty dz\
 {4\over \eta^2(iz)}\ \sum_{j=1}^{N} \Gamma_j^2\ \chi_j(iz).}
Similarly one obtains for the cylinder amplitude
\eqn\cylto{ \wt{A}= \int_0^\infty dz\ {1\over 4\eta^2(iz)}\
 \sum_{j=1}^{N} \left[\sum_a B_a^j\ n^a\right]^2\ \chi_j(iz)}
and for the M\"obius amplitude
\eqn\moebiust{ \wt{M}=\int_0^\infty dz {2\over \hat\eta^2\left({1\over 2}+
 iz\right)}\
 \sum_j \Gamma_j\ \left[\sum_a B_a^j\ n^a\right]\
 \hat{\chi}_j\left( {1\over 2}+iz\right). }
Tadpole cancellation means that the overall coefficients of the massless
orbits $M$ in the sum of these three terms $\wt{K}+\wt{A}+\wt{M}$ must
vanish. This usually yields a highly non-trivial set of linear equations
for the positive integer valued CP factors. In contrast to the advanced model
building machinery in
heterotic or \typea\ string theory, \type models really need
a thorough case by case study. After determining the CP factors,
the closed and open
string spectrum is determined by the projections ${1\over 2}(Z_T+K)$ and
${1\over 2}(A+M)$, respectively.

\newsec{\Type on $T^6/\ZZ_3$ }

In reference \rgepner\ Gepner models for six dimensional vacua were
considered, where special
emphasis was put in models corresponding to toroidal orbifolds
decribed by free $N=2$ super conformal field theories.
The main problem one faces in extending such an analysis to four dimensions,
is the huge number of spectral flow invariant orbits $\chi_i$, which is of
order $10^3$. One way to reduce this number
to smaller size (order $10^2$) is to first divide by a further
simple current of integer conformal dimension and order $K$.
This introduces new projections leading to a theory with order $10^3/K^2$
independent characters. In general a simple current might have fixed points,
which means orbits of shorter length, so that one first has to resolve these
fix points in order to define a bona fide conformal field theory \rfss.
For instance, if one of the levels of the $N=2$ tensor factors is even,
fix point resolution becomes already inevitable.
In the following some of the very few cases without fix points are considered.

As a first warm off example we consider the Gepner model $(1)^9$ containing
2187 orbits of integer charge \rgepner. Further projection with the two
integer dimensional simple currents
\eqn\simplea{\eqalign{ J_1&=\Phi^0_{0,0}\otimes \Phi^0_{0,0}\otimes
 \Phi^0_{0,0}\otimes
 \Phi^1_{-1,0}\otimes \Phi^1_{-1,0}\otimes
 \Phi^1_{-1,0}\otimes \Phi^1_{1,0}\otimes
 \Phi^1_{1,0}\otimes \Phi^1_{1,0} \cr
 J_2&=\Phi^0_{0,0}\otimes \Phi^1_{-1,0}\otimes
 \Phi^1_{1,0}\otimes
 \Phi^0_{0,0}\otimes \Phi^1_{-1,0}\otimes
 \Phi^1_{1,0}\otimes \Phi^0_{0,0}\otimes
 \Phi^1_{-1,0}\otimes \Phi^1_{1,0} \cr}}
reduces the number of orbits to 27.
The charge conjugation modular invariant leads to a \typeb model with
$36$ hypermultiplets and no vectormultiplet.
This is the same spectrum as for the $\ZZ_3$ orbifold of the six torus
$T^6$ with Hodge numbers $h_{11}=36$ and $h_{21}=0$.
It was shown in \rfourdim\ that the geometric \type orientifold has maximal
gauge group $U(12)\times SO(8)$.
We will see that in the Gepner model framework we do not come even close.
Both the charge conjugation and the diagonal torus partition function
yield \type descendants with $36$ neutral chiral multiplets in the closed
string sector and with maximal gauge group $Sp(4)$ in the open string sector.
For the diagonal invariant this spectrum is enlarged by $36$ further
chiral multiplets in the adjoint of $Sp(4)$. The result for the
charge conjugation invariant is consistent
with the prediction that a backgound two form field of rank $r=6$ reduces the
rank of the gauge group by a factor $2^{r/2}$ \refs{\rkst,\rbback}.
Thus, as this model seems to suggest, Gepner models might lead only to \type
descendants which lie on some fairly uninteresting branches
of the moduli space. Nevertheless, for non-toroidal Calabi-Yau
manifolds we do not have any better technique available and we will see
in the next section that more interesting gauge groups can arise.

\newsec{\Type on an orbifold of the Quintic}

Consider the \typeb compactification on the quintic $\IP_{4}[5]$.
At the Landau-Ginzburg point in K\"ahler moduli space it is decribed
by the conformal field with tensor product $(k=3)^5$ in the internal
sector. By taking orbits with respect to the GSO projections one obtains
$n=4000$ characters, $202$ of them are massless and for the diagonal
invariant give rise to the well known
spectrum of $h_{21}=101$ vector multiplets and $h_{11}=1$ hypermultiplets.
We use the freely acting simple current
\eqn\simpleb{ J=\Phi^0_{0,0}\otimes \Phi^3_{-3,0}\otimes \Phi^3_{-1,0}
 \otimes \Phi^3_{1,0} \otimes \Phi^3_{3,0} }
to reduce the number of characters to $n=160$. Now, only $42$ of them
are massless and the diagonal \typeb spectrum consists of $h_{21}=21$ vector
multiplets and $h_{11}=1$ hypermultiplets. This is equivalent to modding
out the quintic by the discrete $\ZZ_5$ subgroup
$z_m\to e^{2\pi i{m\over 5}}z_m$.

\subsec{The charge conjugation invariant}

Calculating the matrices $S$, $C$, $T$ and $P$ for these $160$ characters,
it turns out that there exist $32$ self-conjugate orbits $O_c$. The massless
orbits imply $42$ tadpole cancellation conditions for the $160$
CP factors. These conditions are linear in the CPs, but of course
have to be solved over non-negative integers.
Even the fastest computer can not check all possibilities.
One solution is obvious from the fact, that there exists an orbit $\Phi$
with fusion rules
\eqn\fus{ \Phi\times \Phi=\sum_{i\in O_c} \Phi_i.}
Thus, there exists a CP factor $n$ such that $A=n^2\, K$ and one obtains
only one independent tadpole cancellation condition, which is solved by $n=4$.
The massless spectrum consists of the usual $N=1$ supergravity multiplet,
a chiral multiplet containing the dilaton, $22$ further neutral
chiral multiplets all appearing in the projected closed string sector.
In the open sector one finds a vector multiplet in the adjoint of
$SO(4)$ and in each case one chiral multiplet in the symmetric,
{\bf 9}, and in the singlet, {\bf 1}, representation of $SO(4)$.

Similar solutions with gauge group $SO(4)$ exist in all the other examples
studied in this paper, as well. Note, that the reduction of the open string
gauge group to $SO(4)$ or $Sp(4)$ is what one would naively expect for a
toroidal orbifold with an NS two form background of maximal rank \rbback.

\subsec{The diagonal invariant}

Instead of using the charge conjugate torus partition function we
can also use the diagonal one. Then, the Klein bottle amplitude contains
all $160$ orbits. In order to find the CP assignment in the cylinder amplitude
via \kbrel\ and \crel\
we have to know a simple current relating the charge conjugate
invariant to the diagonal invariant. This is easy to find and is given by
\eqn\simpleb{ J=\Phi^0_{0,0}\otimes \Phi^3_{1,0}\otimes \Phi^3_{-1,0}
 \otimes\Phi^3_{-1,0} \otimes \Phi^3_{1,0}. }
Identifying all CPs in the charge conjugate annulus amplitude lying in
the same orbits under \simpleb\ yields $32$ CP factors for the
diagonal invariant. Since there are now only two(!) non-trivial tadpole
cancellation conditions for $32$ variables we can solve the equations
in full generality, revealing a vast amount of different solutions.
The largest gauge group one can get is $SO(12)\times SO(20)$ without any
charged matter. The massless spectrum from the closed sector is the same
as for the charge conjugation invariant.
A few other solutions of the tadpole cancellation condition are presented
in Table 1. Since one expects non-trivial superpotentials to arise,
a more detailed investigation is needed to decide which models
lie on the same moduli.
\bigno
\cl{\vbox{
\hbox{\vbox{\offinterlineskip
\def\tablespace{height2pt&\omit&&\omit&&\omit&&
 \omit&\cr}
\def\tablerule{\tablespace\noalign{\hrule}\tablespace}

\hrule\halign{&\vrule#&\strut\hskip0.2cm\hfil#\hfill\hskip0.2cm\cr
\tablespace
& Invariant && closed sector && gauge group && matter &\cr
\tablerule
& $C$ && $22\times {\bf 1}_L$ && $SO(4)$ && ${\bf 9}_L$ &\cr
& && && && ${\bf 1}_L$ &\cr
\tablerule
& $D$ && $22\times {\bf 1}_L$ && $SO(12)\times SO(20)$ && -- &\cr
\tablerule
& $D$ && $22\times {\bf 1}_L$ && $SO(12)\times SO(20-n)\times SO(n)$ &&
 $ ({\bf 1},{\bf 20-n},{\bf n})_L$ &\cr
\tablerule
& $D$ && $22\times {\bf 1}_L$ && $SO(10)\times SO(18)\times U(1)$ &&
 $ ({\bf 10},{\bf 1};\pm 1)_L$ &\cr
& && && && $ ({\bf 1},{\bf 18};\pm 1)_L$ &\cr
& && && && $ ({\bf 1},{\bf 1};0)_L$ &\cr
\tablerule
& $D$ && $22\times {\bf 1}_L$ && $SO(6)\times SO(8)\times SO(6)$ &&
 $ ({\bf 6},{\bf 1},{\bf 6})_L$ &\cr
& && && && $2\times ({\bf 1},{\bf 8},{\bf 6})_L$ &\cr
& && && && $3\times ({\bf 1},{\bf 1},{\bf 15})_L$ &\cr
& && && && $({\bf 1},{\bf 1},{\bf 20})_L$ &\cr
& && && && $({\bf 1},{\bf 1},{\bf 1})_L$ &\cr
\tablespace}\hrule}}}}
\cl{
\hbox{{\bf Table 1:}{\it ~~massless spectra for \type quintic descendant }}}

Unfortunately, in the rational CFT setting we can not
decide whether, for instance, the $SO(12)\times SO(20)$ model contains
any 5-branes or is perturbative from the heterotic viewpoint. Finding
a possible heterotic dual would be interesting but is beyond the scope
of this paper.
Note, that all models in Table 1 are non-chiral and
as expected from one-loop consistency all gauge anomalies vanish.
We would like to emphasize, that the geometric reason for getting so
many solutions is that
the number of K\"ahler moduli and therefore the number of tadpole cancellation
conditions for this peculiar Calabi-Yau is so small, namely
$h_{11}=1$. Thus, in order to see similar patterns in other
Gepner models, one should in particular study cases with small numbers
of K\"ahler moduli and interacting $N=2$ superconformal field theories.

\newsec{\Type on an orbifold of $\IP_{1,1,1,3,3}[9]$}
The second model we want to study is an orbifold of the Calabi-Yau
$\IP_{1,1,1,3,3}[9]$, which has Hodge numbers $h_{21}=104$ and $h_{11}=4$.
The model itself contains the solvable Gepner point $(1^2\, 7^3)$ and
gives rise to $5184$ orbits after imposing the GSO
projection. Reducing this number to $576$ is possible by also projecting
with the simple current
\eqn\simpleb{ J=\Phi^0_{0,0}\otimes \Phi^0_{0,0}\otimes \Phi^7_{-3,0}\otimes
 \Phi^7_{3,0} \otimes \Phi^0_{0,0} }
of order $3$. The Hodge numbers of the orbifold theory are $h_{21}=46$ and
$h_{11}=10$.
The calculation is similar to the one for the quintic.
For the charge conjugation invariant we obtain 67 tadpole cancellation
conditions, two solutions of which are displayed in Table 2. In this case,
we did not find any solution with gauge groups bigger than products of
$SO(4)$ factors.
\bigno
\cl{\vbox{
\hbox{\vbox{\offinterlineskip
\def\tablespace{height2pt&\omit&&\omit&&\omit&&
 \omit&\cr}
\def\tablerule{\tablespace\noalign{\hrule}\tablespace}

\hrule\halign{&\vrule#&\strut\hskip0.2cm\hfil#\hfill\hskip0.2cm\cr
\tablespace
& Invariant && closed sector && gauge group && matter &\cr
\tablerule
& $C$ && $56\times {\bf 1}_L$ &&
 $SO(4)$ && $7\times {\bf 9}_L$ &\cr
& && && && $7\times {\bf 1}_L$ &\cr
& && && && $1\times {\bf 6}_L$ &\cr
\tablerule
& $C$ && $56\times {\bf 1}_L$ && $SO(4)^3$ &&
$2\times ({\bf 9},{\bf 1},{\bf 1})_L$ &\cr
& && && && $2\times ({\bf 1},{\bf 1},{\bf 1})_L$ &\cr
& && && && $ ({\bf 4},{\bf 4},{\bf 1})_L$ &\cr
& && && && $ ({\bf 1},{\bf 4},{\bf 4})_L$ &\cr
& && && && $ ({\bf 4},{\bf 1},{\bf 4})_L$ &\cr
\tablerule
& $D$ && $56\times {\bf 1}_L$ && $SO(8)^4\times SO(4)^3$ &&
$2\times ({\bf 8},{\bf 8},{\bf 1},{\bf 1};{\bf 1},{\bf 1},{\bf 1})_L$ &\cr
& && && &&
$2\times ({\bf 8},{\bf 1},{\bf 8},{\bf 1};{\bf 1},{\bf 1},{\bf 1})_L$ &\cr
& && && &&
$2\times ({\bf 8},{\bf 1},{\bf 1},{\bf 8};{\bf 1},{\bf 1},{\bf 1})_L$ &\cr
& && && &&
$1\times ({\bf 8},{\bf 1},{\bf 1},{\bf 1};{\bf 4},{\bf 1},{\bf 1})_L$ &\cr
& && && &&
$1\times ({\bf 8},{\bf 1},{\bf 1},{\bf 1};{\bf 1},{\bf 4},{\bf 1})_L$ &\cr
& && && &&
$1\times ({\bf 8},{\bf 1},{\bf 1},{\bf 1};{\bf 1},{\bf 1},{\bf 4})_L$ &\cr
& && && &&
$1\times ({\bf 1},{\bf 8},{\bf 8},{\bf 1};{\bf 1},{\bf 1},{\bf 1})_L$ &\cr
& && && &&
$1\times ({\bf 1},{\bf 8},{\bf 1},{\bf 8};{\bf 1},{\bf 1},{\bf 1})_L$ &\cr
& && && &&
$2\times ({\bf 1},{\bf 8},{\bf 1},{\bf 1};{\bf 4},{\bf 1},{\bf 1})_L$ &\cr
& && && &&
$1\times ({\bf 1},{\bf 8},{\bf 1},{\bf 1};{\bf 1},{\bf 4},{\bf 1})_L$ &\cr
& && && &&
$2\times ({\bf 1},{\bf 8},{\bf 1},{\bf 1};{\bf 1},{\bf 1},{\bf 4})_L$ &\cr
& && && &&
$1\times ({\bf 1},{\bf 1},{\bf 8},{\bf 8};{\bf 1},{\bf 1},{\bf 1})_L$ &\cr
& && && &&
$2\times ({\bf 1},{\bf 1},{\bf 8},{\bf 1};{\bf 4},{\bf 1},{\bf 1})_L$ &\cr
& && && &&
$2\times ({\bf 1},{\bf 1},{\bf 8},{\bf 1};{\bf 1},{\bf 4},{\bf 1})_L$ &\cr
& && && &&
$1\times ({\bf 1},{\bf 1},{\bf 8},{\bf 1};{\bf 1},{\bf 1},{\bf 4})_L$ &\cr
& && && &&
$1\times ({\bf 1},{\bf 1},{\bf 1},{\bf 8};{\bf 4},{\bf 1},{\bf 1})_L$ &\cr
& && && &&
$2\times ({\bf 1},{\bf 1},{\bf 1},{\bf 8};{\bf 1},{\bf 4},{\bf 1})_L$ &\cr
& && && &&
$2\times ({\bf 1},{\bf 1},{\bf 1},{\bf 8};{\bf 1},{\bf 1},{\bf 4})_L$ &\cr
& && && &&
$1\times ({\bf 1},{\bf 1},{\bf 1},{\bf 1};{\bf 4},{\bf 4},{\bf 1})_L$ &\cr
& && && &&
$1\times ({\bf 1},{\bf 1},{\bf 1},{\bf 1};{\bf 4},{\bf 1},{\bf 4})_L$ &\cr
& && && &&
$1\times ({\bf 1},{\bf 1},{\bf 1},{\bf 1};{\bf 1},{\bf 4},{\bf 4})_L$ &\cr
\tablespace}\hrule}}}}
\cl{
\hbox{{\bf Table 2:}{\it ~~massless spectra for \type descendant of
$(1)^2(7)^3$ Gepner model}}}

The diagonal invariant leads to only nine conditions for $64$ CP factors.
A lot of solutions can be found by computer search, however, there is no
chance of classifying them all. In Table 2 we present the solution we found
with maximal rank of the gauge group. Remarkably, the rank is $22$ so that
we can conclude that there must be 5-branes present in this vacuum and it
must correspond to a non-perturbative vacuum on the dual heterotic side.
The lesson we learn from the latter two examples is, that in contrast to
some expectations Gepner models can provide four dimensional, $N=1$
supersymmetric \type backgrounds with interesting gauge groups.

\newsec{Summary}
In this paper we have started an investigation of four dimensional
\type vacua with $N=1$ supersymmetry derived from Gepner models.
Geometrically, these vacua correspond to compactifications on Calabi-Yau
manifolds. We have computed only the simplest non-toroidal models, but
we think the results are encouraging for further study in this rich
class of models. To get interesting results, one should in particular
consider models with small numbers of K\"ahler moduli.
Technically, there exist some generalizations to the constuction
presented in this paper like the choice of non-trivial Klein Bottle
projections or the choice of non-trivial gluing automorphisms for
the cylinder amplitude.

With respect to the question
of heterotic dual models, it would be very nice to have an intrinsic
condition to distinguish \type backgrounds with and without 5-branes.
Furthermore, the existence of $(0,2)$ target space dualities for perturbative
heterotic vacua \rbsf\ suggests that such 
dualities should also hold for at least a subclass of \type vacua.

\bigskip\bigskip\centerline{{\bf Acknowledgments}}\pano
It is a pleasure to thank Angel Uranga and Jaemo Park for discussion and
Zurab Kakushadse for valuable e-mail correspondence. Moreover, we would like
to thank J\"urgen Fuchs, Christoph Schweigert and Andreas Recknagel
for some comments about an earlier version of this paper.
This work is supported by NSF grant PHY--9513835.
\listrefs
\bye